\documentclass[amssymb,aps]{revtex4}
\usepackage{amsmath}
\usepackage{amsfonts,amssymb,amsthm, mathrsfs}
\usepackage[utf8]{inputenc}
\usepackage{graphicx}
\usepackage{hyperref}
\usepackage[usenames,dvipsnames]{color}
\begin{document}
\title{Topological Model for Domain Walls in (Super-)Yang-Mills Theories}
\author{Markus Dierigl}
\email{markus.dierigl@mytum.de}
\affiliation{Arnold-Sommerfeld-Center, Ludwig-Maximilians-Universit\"at, Theresienstr. 37, 80333 M\"unchen}
\affiliation{Technische Universit\"at M\"unchen, James-Franck-Str. 1, 85748 Garching}
\author{Alexander Pritzel}
\email{alexander.pritzel@physik.lmu.de}
\affiliation{Arnold-Sommerfeld-Center, Ludwig-Maximilians-Universit\"at, Theresienstr. 37, 80333 M\"unchen}
\date{\today}

\begin{abstract}
We derive a topological action that describes the confining phase of (Super-)Yang-Mills theories with gauge group $SU(N)$, similar to the work recently carried out by Seiberg and collaborators. It encodes all the Aharonov-Bohm phases of the possible non-local operators and phases generated by the intersection of flux tubes. Within this topological framework we show that the worldvolume theory of domain walls contains a Chern-Simons term at level $N$ also seen in string theory constructions.\\
The discussion can also illuminate dynamical differences of domain walls in the supersymmetric and non-supersymmetric framework. Two further analogies, to string theory and the fractional quantum Hall effect might lead to additional possibilities to investigate the dynamics.
\end{abstract}

\maketitle

\section{Introduction}

Domain Walls in Super-Yang-Mills theories have been under intensive investigation since their discovery by Dvali and Shifman in 1996 \cite{Dvali:1996xe}. Many different methods were applied in order to receive information about the properties of the wall configurations.\\
Supersymmetry allows to calculate the wall tension exactly under the assumption that the walls are BPS saturated. In the large $N$ limit for the pure $SU(N)$ gauge theory this calculation suggests that the wall tension is proportional to $N$ rather than $N^2$ as would be expected for solitonic solutions, see \cite{Witten:1997ep}. This functional dependence rather resembles that of D-branes in string theory. In fact, the string theoretical description of these domain walls, \cite{Witten:1997ep}, confirms this dependence and predicts that chromoelectric flux tubes, the analog of open strings, can end on the domain wall. This phenomenon further arises naturally in the Dvali Shifman mechanism of dynamical compactification, \cite{Dvali:1996bg}. Moreover, string theory suggests the occurrence of a level $N$ Chern-Simons term on the wall worldvolume as shown by Acharya and Vafa in \cite{Acharya:2001dz}.\\
Despite the amount of work done, with field theoretical considerations this Chern-Simons term has not been constructed explicitly, although the considerations in \cite{Gaiotto:2013gwa} render such a term plausible as well in a field theoretical context.\\
In this paper we will construct a topological model that describes the topological properties of the confining phase of (Super-)Yang-Mills theories, especially the Aharonov-Bohm phases of non-local operators due to their center charges and phases generated due to a crossing of electric flux tubes. In this topological model a Chern-Simons term at level $N$ arises in the presence of domain walls. Furthermore, chromoelectric flux tubes, described by electric surface operators, can end on the domain walls. First, we work in the framework of non-supersymmetric pure Yang-Mills theory and extend the model to include the supersymmetric case.\\
The paper is organized as follows. In section \ref{YMDW} we discuss the vacuum structure of pure Yang-Mills theories and the wall solutions that are thought to appear in the large $N$ limit. We explicitly derive a topological theory of the confining phase that contains all the exchange phases in dependence of the vacuum angle $\theta$. The introduction of domain walls in this framework immediately leads to a Chern-Simons term at level $N$ due to gauge invariance properties. In section \ref{SYMDW} we extend the considerations to a supersymmetric theory and the original domain walls in Super-Yang-Mills models and describe some implications for the dynamics of the walls themselves in section \ref{dynamics}. In section \ref{analogies} analogies to string theory and the fractional quantum Hall effect are pointed out and hints on how to proceed the investigation are presented. The last section \ref{outlook}  summarizes the results.
\section{Yang-Mills Domain Walls}
\label{YMDW}
First, we discuss the domain walls in pure Yang-Mills theories with gauge group $SU(N)$. The Lagrangian density with 't~Hooft coupling $\lambda \equiv g^2 N$ reads
\begin{equation}
\mathcal{L} = -\frac{N}{2 \lambda} \text{Tr}(F \wedge \ast F) + \frac{\theta}{8 \pi^2} \text{Tr}(F \wedge F) \enspace,
\end{equation}
with non-Abelian field strength tensor $F$. Yang-Mills theory develops a mass gap for low energies and exhibits confinement. In the latter we always assume confinement to be the effect of the condensation of charge $N$ monopoles. This mechanism is well established and is justified by considerations, e.g.\ as presented in \cite{Seiberg:1994rs}.

\subsection{Vacuum structure in the large N limit}

\begin{figure}[htp]
\centering
\includegraphics[width=0.4\textwidth]{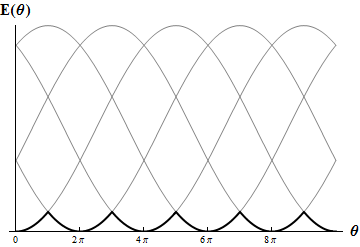}
\caption{$\theta$-dependence of the energy for the $N$ branches}
\label{thetavacua}
\end{figure}

The usual procedure for the large $N$ limit states that the parameter to be kept fixed is $\theta / N$ rather than $\theta$ itself. This, however, would destroy the $2 \pi$-periodicity in $\theta$. Moreover, the $\theta$-dependence of the energy would vanish, being a smooth function of the parameter $\theta / N$. A solution was suggested by Witten in \cite{Witten:1978bc}. He introduces $N$ branches representing quasi-stable states in the large $N$ limit, labeled by an integer parameter $k \in \{0, \dots, N-1\}$. The energy of each branch is $2 \pi N$-periodic in $\theta$, but the whole collection restores the expected $2 \pi$-periodicity and $\theta$-dependence of the energy, see figure \ref{thetavacua} for a schematical picture.
For a shift in $\theta$ the branches interchange roles, i.e.\ for $\theta = 2 \pi$ the branch with lowest energy is labeled by $k = 1$ rather than $k = 0$ as for $\theta = 0$. By the Witten effect, \cite{Witten:1979ey}, we infer that the electric charge of the condensate differs in the quasi-stable configurations. The lowest energy state for the $SU(N)$ theory is always assumed to be a state with purely magnetically charged condensate. This assumption is supported by the considerations in section \ref{dynamics}. Hence, the energetically higher configurations for fixed $\theta$ describe dyonic condensates. Note that $\theta$ and $k$ are distinct parameters of the theory but produce equivalent effects.\\
The reason why these states are quasi-stable is that neighboring vacua show an energy difference that is $\mathcal{O}(N^0)$, whereas the domain walls have a tension $\mathcal{O}(N)$, see \cite{Witten:1998uka}. Thus, vacuum transitions are exponentially suppressed in the large $N$ limit, see as well \cite{Shifman:1998if}.
\subsection{Classification of the vacua}
For the condensation of charge $N$ monopoles as confining mechanism, it is sufficient to classify non-local operators by their electric and magnetic charges and fluxes under the center of the gauge group. For the gauge group $SU(N)$ this leads to a $\mathbb{Z}_N \times \mathbb{Z}_N$ lattice as discussed in \cite{Aharony:2013hda} and originating from \cite{'tHooft:1979uj, 'tHooft:1981ht, Donagi:1995cf}.\\
The different vacua can be categorized by the choice of genuine line operators, discussed in \cite{Aharony:2013hda} and \cite{Kapustin:2014gua}. These are the line operators which do not have to be supplemented by a surface operator. A modified Dirac quantization for line operators of electric and magnetic charge $(q,m)$ (see \cite{Gaiotto:2010be})
\begin{equation}
q m' - m q' \in \mathbb{Z} \enspace,
\end{equation}
has to be fulfilled for these genuine line operators. Note that for convenience we have rescaled the electric charges by a factor of $1/N$ in order to relate them to the center charges fundamental quarks in the theory would have. The line operators that do not fulfill this quantization have to be supplemented by a surface operator, which generates a non-trivial phase (under exchanges with non-vanishing winding). These surface operators can be interpreted as flux tubes connecting the confined charges.\\
There are several possibilities for the choice of the genuine operators described in \cite{Aharony:2013hda}. For a $SU(N)/ \mathbb{Z}_N$ theory the vacuum angle has a $2 \pi N$-periodicity (similar to a single branch in figure \ref{thetavacua}). For fixed $\theta$, there are $N$ possibilities for the charge of the genuine line operators. Because we expect the genuine line operators to be in line with the condensate this can be understood as the choice of the condensate charge, parametrized by $(q,m) = (k/N, N)$ with $k \in \{0, \dots, N-1\}$. This means that there are $N$ distinct theories, labeled by $k$. Nevertheless, in the topological approach (disregarding the $\theta$-dependence of the energy) we can scan the whole set of theories by varying $\theta/2\pi \in \{0, \dots, N-1\}$. Accordingly, by the Witten effect this is equivalent to alter the charge of the condensate. Note that this is only valid in the topological theory where the dynamical properties are not taken into account.\\
In the $SU(N)$ theory, however, we have to include all the branches. The genuine line operators should not alter under the shift $\theta \rightarrow \theta + 2\pi$ and thus are the purely electrically charged Wilson loops. Since now the branches are part of the same theory there is the possibility of the aforementioned domain walls, interpolating between branches for fixed $\theta$.\\
In order to construct our topological model we first consider a $SU(N)/\mathbb{Z}_N$ theory, equivalent to a single branch of the full $SU(N)$ theory. After the derivation of a consistent topological action that includes the Witten effect we add the other branches in order to obtain the topological version of the $SU(N)$ theory.

\subsection{Derivation of the topological action}

\begin{figure}[htp]
\centering
\includegraphics[width=0.3\textwidth]{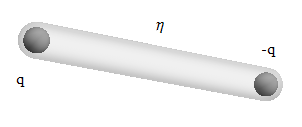}
\caption{Flux tube connecting two probe charges}
\label{fluxtube}
\end{figure}

We start in the theory where charge $N$ monopoles condense to create confinement. This, combined with the classification of the charges via the discrete group $\mathbb{Z}_N$, enables us to describe the topological properties of the $SU(N)$ gauge theory in a topological field theory framework. The discrete gauge group $\mathbb{Z}_N$ can be formulated in analogy to the Abelian-Higgs model by the condensation of charge $N$ monopoles, see for example \cite{Banks:2010zn} and \cite{Gukov:2013zka}. Far in the Higgs regime the action becomes a mere constraint equation. With the dual gauge field $\tilde{A}$, the phase of the magnetic scalar potential $\varphi$ with charge $N$, and the Lagrange multiplier 3-form $h$ the Euclidean action reads (see \cite{Gukov:2013zka})
\begin{equation}
S = \int{ h \wedge (d \varphi - N \tilde{A})} \enspace,
\end{equation}
with 0-form gauge transformations ($f$ a $2\pi$-periodic 0-form)
\begin{equation}
\begin{split}
\varphi & \rightarrow \varphi + N f,\\
\tilde{A} & \rightarrow \tilde{A} + d f \enspace.
\end{split}
\end{equation}
Dualizing the field $\tilde{A}$ in order to retrieve the usual gauge field $A$ and the scalar field $\varphi$ to a 2-form field $B$ coupling to electric flux tubes we obtain
\begin{equation}
\begin{split}
S =& \int{\left[h \wedge (d \varphi - N \tilde{A}) + \frac{i}{2\pi} d \varphi \wedge d B + \frac{i}{2\pi} d \tilde{A} \wedge d A\right]}\\
=& \frac{i}{2\pi} \int{ \tilde{F} \wedge (F - N B)} \enspace.
\end{split}
\label{dscaction}
\end{equation}
This has the functional form of a BF-theory introduced in \cite{Horowitz:1989ng}, a topological field theory in four dimensions which encodes the exchange phases for line and surface operators.\\
The action has an additional 1-form gauge symmetry. With 1-form $\lambda$ which fulfills the quantization condition over a closed 2-surface
\begin{equation}
\frac{1}{2\pi} \oint d \lambda \in \mathbb{Z} \enspace,
\end{equation}
it reads
\begin{equation}
\begin{split}
A & \rightarrow A + N \lambda, \\
B & \rightarrow B + d \lambda \enspace.
\end{split}
\end{equation}
This action describes the topological pro\-per\-ties of char\-ges and fluxes in a dual superconductor at $\theta = 0$. All the 't~Hooft loops of magnetic charge $m$ 
\begin{equation}
\mathcal{H}(m) = \exp{\left(i m \oint{\tilde{A}} \right)} \enspace,
\end{equation}
are gauge invariant. The Wilson loops of charge $q$ along a path $\gamma$ have to be supplemented by a surface operator of electric flux $\eta$ over an open surface $\Sigma$, with $\partial \Sigma = \gamma$
\begin{equation}
\mathcal{W}(q, \eta) = \exp{\left( i q \oint_{\gamma}{A} - i \eta N \int_{\Sigma} B \right)} \enspace.
\end{equation}
With Stokes' theorem this transforms as
\begin{equation}
\Delta \mathcal{W}(q, \eta) = \exp{\left(i q N \oint_{\gamma}{\lambda} - i \eta N \oint_{\partial \Sigma}{\lambda}\right)} \enspace,
\end{equation}
and $\Delta \mathcal{W}$ vanishes for $q = \eta$, as desired. This objects can be understood as two electric probe charges connected by a flux tube, see figure \ref{fluxtube}.
The action \eqref{dscaction} with the corresponding 1-form gauge transformations hence successfully describes the topological properties of a dual superconductor. The $\theta$-angle and the corresponding Witten effect, however, are not yet incorporated.\\
After a shift of $\theta$ by $2\pi$ the condensate develops electric charge and instead of pure 't~Hooft loops we expect dyonic loops to be gauge invariant on their own. This can be achieved by modifying the 1-form gauge transformations of the dual gauge field $\tilde{A}$
\begin{equation}
\tilde{A} \rightarrow \tilde{A} - \frac{\theta}{2\pi} \lambda \enspace.
\end{equation}
Nevertheless, the action does not remain gauge invariant under this transformations
\begin{equation}
\Delta S = -\frac{i \theta}{4\pi^2} \int{ d \lambda \wedge (F -NB)} \enspace.
\end{equation}
The term proportional to $d \lambda \wedge F$ is of no concern, because for $\theta \in 2\pi \mathbb{Z}$ its integrated value is a multiple of $2\pi$ and the action remains unaltered. The second term has to be compensated by an additional contribution to the action. In four dimensions a $B \wedge B$-term is allowed, see \cite{Horowitz:1989ng}, and this indeed has the appropriate transformation properties. Thus, the generalized action including the Witten effect reads
\begin{equation}
S = \frac{i}{2\pi} \int{\left[ \tilde{F} \wedge (F - N B) - \frac{N \theta}{4 \pi} B \wedge B \right]} \enspace,
\end{equation}
with the 1-form gauge transformations
\begin{equation}
\begin{split}
B & \rightarrow B + d \lambda,\\
A & \rightarrow A + N \lambda,\\
\tilde{A} & \rightarrow \tilde{A} - \frac{\theta}{2\pi} \lambda \enspace.
\end{split}
\label{gaugetrafo}
\end{equation}
The $B \wedge B$-term accounts for the phase created for crossing flux tubes. Integrating out $\tilde{F}$, we see that the $B \wedge B$-term describes a $F \wedge F$ contribution in the topological action, \cite{Kapustin:2014gua}
\begin{equation}
S \rightarrow \frac{i}{8 \pi^2 N} \int{ \theta F \wedge F} \enspace.
\end{equation}
The prefactor $1/N$ further suggests that the intersections of strings behave as fractional ($1/N$) instantons and further account for the $2 \pi N$-periodicity of $\theta$ for gauge group $SU(N)/\mathbb{Z}_N$. The strong relation between the $B \wedge B$- and $F \wedge F$-term can be illustrated in the following way. Consider a resting electric flux tube (e.g.\ along $\hat{z}$-direction) that is crossed by another electric flux tube (e.g.\ oriented in $\hat{x}$- and moving in $\hat{y}$-direction). The moving flux tube induces a magnetic field (in the $\hat{z}$-direction) and therefore leads to a contribution of the form $\vec{E} \cdot \vec{B} \propto F \wedge F$, see \cite{Preskill:2000}.\\
Now we include the $N$ branches to recover the $SU(N)$ theory, which has to our knowledge not been done in previous investigations, we introduce the integer valued parameter $k$ labeling the branch of the $SU(N)$ theory. The correspondence of the branches and the charges of the condensate dictate its form and the full action including Witten effect and the set of branches is
\begin{equation}
S = \frac{i}{2\pi} \int{\left[ \tilde{F} \wedge (F - N B) - \frac{N \theta}{4 \pi} B \wedge B  + \frac{N k}{2} B \wedge B \right]} \enspace.
\end{equation}
The $2 \pi$-periodicity is restored and the only valid choice for the genuine line operators are the Wilson loops (consider a shift $\theta \rightarrow \theta +2 \pi$). To preserve the gauge invariance of the action the dual gauge field transforms as
\begin{equation}
\tilde{A} \rightarrow \tilde{A} - \frac{\theta}{2 \pi} \lambda + k \lambda \enspace.
\end{equation}
A similar action was presented in \cite{Kapustin:2014gua} without explicit derivation. There the authors coupled this topological theory to a dynamical $SU(N)$ theory in order to obtain a dynamical $SU(N)/\mathbb{Z}_N$ theory.

\subsection{Domain walls and the Chern-Simons term}

Domain walls in pure Yang-Mills theories with gauge group $SU(N)$ interpolate between quasi-stable configurations labeled by the integer $k$ at fixed $\theta$. The difference in the energy density renders them non-static which, however, is of no concern for the topological discussion.\\
In the non-supersymmetric framework we can set $\theta$ to zero in the following analysis. Thus, fundamental domain walls, with $\Delta k = 1$, are encoded via the jump of $k$ on a codimension one surface $\mathcal{V}$. This has direct consequences for the transformation properties of the action
\begin{equation}
\begin{split}
\Delta S =& \frac{i}{2\pi} \int{ d\left[k \lambda \wedge F + \frac{Nk}{2} \lambda \wedge d \lambda \right]} \\
& - \frac{iN}{4\pi} \int{ d k \wedge (2 \lambda \wedge B + \lambda \wedge d \lambda)} \enspace.
\end{split}
\end{equation}
The first term is a total derivative and only contributes for boundaries of the spacetime manifold, which we do not consider here. The second term, however, develops a contribution on the domain wall worldvolume
\begin{equation}
\Delta S_{\text{wall}} = -\frac{iN}{4\pi} \int_{\mathcal{V}}{(2 \lambda \wedge B + \lambda \wedge d \lambda)} \enspace.
\label{deltawallaction}
\end{equation}
To allow for domain walls in the topological theory and simultaneously preserve the gauge invariance we have to introduce a new boundary field that transforms under the 1-form gauge transformations. The natural choice is a 1-form field $\mathcal{A}$ with 
\begin{equation}
\mathcal{A} \rightarrow \mathcal{A} - \lambda
\end{equation}
The action gets modified on the worldvolume of the wall and has to include
\begin{equation}
S_{\mathcal{V}} = - \frac{i N}{4 \pi} \int_{\mathcal{V}}{ ( 2 \mathcal{A} \wedge B + \mathcal{A} \wedge d \mathcal{A} )} \enspace.
\label{wallaction}
\end{equation}
The field $\mathcal{A}$ couples to the electric flux tubes and furthermore generates a Chern-Simons term at level $N$ on the domain wall, as was predicted in a string theory approach in \cite{Acharya:2001dz}.

\subsection{Flux tubes ending on domain walls}

\begin{figure}[htp]
\centering
\includegraphics[width=0.2\textwidth]{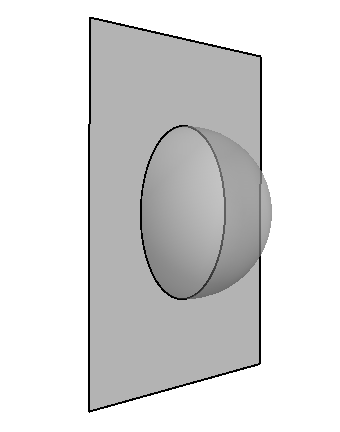}
\caption{Flux tube ending on the domain wall}
\label{wallflux}
\end{figure}

A further phenomenon originating from previous field theoretical considerations and string theory is the claim that electric flux tubes can end on the walls. This can be understood in the topological setup above. Without the presence of domain walls open surface operators have to be bounded by a Wilson loop. With the additional wall fields there is another possibility. By demanding that the boundary of the open surface $\partial \Sigma$ is a subset of the worldvolume $\mathcal{V}$ the gauge invariance can be restored using loop operators of the field $\mathcal{A}$ instead
\begin{equation}
\mathcal{W}_{\text{wall}} = \exp{\left( i N \eta \int_{\Sigma}{B} + i N q \oint_{\partial \Sigma}{\mathcal{A}}\right)} \enspace.
\end{equation}
This enables flux tubes to end and the domain walls, just as open strings end on D-branes, depicted in figure \ref{wallflux}.
In the following, we generalize the topological theory to supersymmetric models.

\section{Super-Yang-Mills Domain Walls}
\label{SYMDW}

\begin{figure}[htp]
\centering
\includegraphics[width=0.3\textwidth]{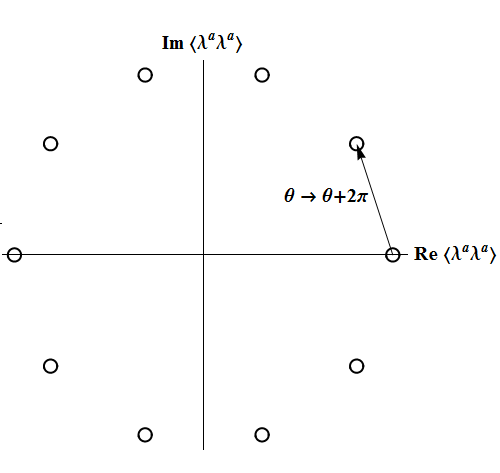}
\caption{Vacua in Super-Yang-Mills theories}
\label{thetacond}
\end{figure}

Super-Yang-Mills theories with gauge group $SU(N)$ differ dramatically in their vacuum structure compared to their non-supersymmetric relatives. In fact, the fermionic superpartners of the gluons, the gluinos $\lambda^a$, condense and one finds $N$ real supersymmetric vacua of vanishing energy density, see \cite{Shifman:1987ia} and \cite{Witten:1982df}, parametrized by $j \in \{0, \dots, N-1\}$
\begin{equation}
\langle \lambda^a \lambda^a \rangle = N \Lambda^3 \exp{\left( 2 \pi i \frac{j}{N} \right)} \enspace,
\end{equation}
with dynamically generated scale $\Lambda$ of the pure gluodynamics.\\These vacua, connected by a non-anomalous chiral $\mathbb{Z}_N$ symmetry, get interchanged as $\theta$ shifts by a multiple of $2\pi$, depicted in figure \ref{thetacond}.
The dynamical phase of the gluino condensate, replacing the integer label $k$, acts as an axionic field rendering $\theta$ dynamical. Similarly, the domain walls can be described as axionic domain walls.\\
For the axionic domain walls the dynamical $\theta$-angle varies by $2\pi$ which is equivalent to a shift in the phase of the gaugino condensate. Consequently, the same topological action can be used in order to encode the exchange properties of non-local operators. Due to the combination of the gluino condensate phase and the $\theta$-parameter, it is sufficient to keep the $\theta$-dependent term, where $\theta$ now is dynamical rather than a fixed parameter (i.e.\ possibility of $d \theta \neq 0$)
\begin{equation}
S = \frac{i}{2 \pi} \int{\left[ \tilde{F} \wedge (F - NB) - \frac{N \theta}{4 \pi} B \wedge B \right]} \enspace,
\end{equation}
with the 1-from gauge transformations \eqref{gaugetrafo}.\\
The action receives a contribution from the jump in $\theta$ of the form
\begin{equation}
\Delta S = \frac{i N}{8 \pi^2} \int{d \theta \wedge (2 \lambda \wedge B + \lambda \wedge d \lambda)} \enspace.
\end{equation}
For a sharp jump of $\theta$ by $2 \pi$ on a codimension one surface $\mathcal{V}$ we once more encounter the wall contribution \eqref{deltawallaction}. Again the boundary field $\mathcal{A}$ has to be introduced with the same Chern-Simons term at level $N$ and coupling to the field $B$ as in \eqref{wallaction}. The mechanism of flux tubes ending on the wall sustains as well.\\
Again, intersections of electric flux tubes generate a phase that resembles a fractional instanton. In Super-Yang-Mills theories this allows a further interpretation.\\
The axionic phase of the gluino condensate corresponds to a pseudoscalar glueball which has a certain overlap with a gaugino bilinear. A single instanton has $2N$ gaugino zero modes, i.e.\ it couples to $2N$ gauginos. A fractional ($1/N$) instanton should therefore couple to two gluinos and have two gaugino zero modes, which is precisely needed to couple to the axionic field. Thus, the $\theta$-term has a factor of $1/N$. Note, however, that the whole theory (no vacuum chosen) remains $2 \pi$-periodic in $\theta$.

\section{Dynamics of the domain walls}
\label{dynamics}

Even though a topological theory does not contain propagating degrees of freedom, some qualitative properties concerning the dynamics of the domain walls themselves can be deduced using some extensions.\\
The $\theta$-term can be rewritten as a total derivative of a 3-form $C$, the Chern-Simons 3-form. In four dimensions a massless 3-form does not contain any propagating degrees of freedom, but supports a long-range electric field via its 4-form field strength $F_4$ in analogy to electrodynamics in two dimensional spacetime. Domain walls couple to $C$ with their three dimensional worldvolume. If we further introduce a kinetic term for the non-propagating 3-form field $C$ of the form
\begin{equation}
d C \wedge \ast (d C) = F_4 \wedge \ast F_4 \enspace,
\end{equation}
the equations of motion in the background of a Yang-Mills domain wall become
\begin{equation}
d (\ast F_4) - \kappa d k = 0 \enspace,
\end{equation}
where $\kappa$ is a non-zero constant. Thus, the electric field jumps to a non-zero value. This term contributes to the energy density as $F_4^2$, leading to the non-degeneracy of the quasi-stable states in the pure Yang-Mills theory and reproducing the $k^2$-dependence predicted in \cite{Witten:1998uka}. This leads to a confinement of a wall-antiwall system similarly to the model in \cite{Luscher:1978rn} by the emergent long-range interaction. That $F_4$ does not vanish for configurations with a dyonic condensate is suggestive as well by rewriting the $F \wedge F$-term in the action as
\begin{equation}
F \wedge F \propto \vec{E} \cdot \vec{B} \enspace.
\end{equation}
Obviously this is zero for a purely magnetically charged condensate but not for dyons.\\
In the supersymmetric extension the picture changes. Now, the phase of the gluino condensate acts as an axionic field and the domain wall can be described as an axionic domain wall. This axion, however, gets eaten up by the 3-form field $C$ as described in \cite{Dvali:2005an}. Both of the fields, axionic gluino condensate phase and massless non-dynamical 3-form $C$, combine to one massive 3-form field which propagates a single degree of freedom. The mass screens the electric field and there is no long-range interaction due to $F_4$ between the walls. If the domain walls in Super-Yang-Mills theory are to be BPS saturated states (see \cite{Dvali:1999pk} and \cite{Gaiotto:2013gwa}), this is necessary because otherwise the configurations would violate the corresponding energy bound.\\
A similar phenomenon occurs for the $CP^N$-model in two dimensions, see \cite{Witten:1978bc}. The confined solitonic configuration become unconfined in the supersymmetric case.

\section{Analogies to other physical models}
\label{analogies}
In order to extract dynamical properties in further investigations, analogies to other physical models might prove useful. Following, we discuss the relation to string theory and to fractional quantum Hall systems.

\subsection{Analogy to string theory}
It is an rather old idea, that Yang-Mills theory in the $1/N$ expansion should behave as a non-critical string theory with $1/N$ controlling the string coupling, \cite{'tHooft:1973jz}. Within this context the domain walls discussed here were identified as close analogs to D-branes in \cite{Witten:1997ep,Witten:1998uka}. It is a very interesting question how far this analogy can be pushed. We try to illustrate that there are some non-trivial implications. \\
Let us recall that in critical type II string theories p-branes always come in pairs of a p-brane and a q-brane, with
\begin{equation}
p+q = 6 \enspace.
\label{pqformula10}
\end{equation}
Here the p-brane carries an electric charge under a massless Ramond-Ramond (p+1)-form, while the corresponding (6-p)-brane carries a magnetic charge, i.e.\ it is described by a winding in the (p+1)-form field. A slightly different situation arises when we consider branes, which carry charges under a finite group, this situation arises for example in type I string theory as discussed in \cite{Witten:1998cd}. Here we again have a pairing between a p-brane and a q-brane, but with
\begin{equation}
	p+q = 7 \enspace.
	\label{pqformula10mod}
\end{equation}
Now, we try to understand whether an analogous situation arises in our setup. In the supersymmetric case, the 3-form coupling to the domain wall acquired a mass by combining with a pseudoscalar glueball (the axionic phase of the gaugino condensate), so we do not have a massless 3-form, furthermore our walls are charged under a chiral $Z_N$ symmetry as discussed earlier. This suggests, that our setup should be closer to the case of branes with discrete charges in critical string theory. This suggests that the object paired up with the domain walls should satisfy a relation, that is the four dimensional analogue of (\ref{pqformula10mod}), i.e.
\begin{equation}
p+q = 1 \enspace.
\end{equation}
For the domain wall case of $p=2$ this suggests we are looking for a D(-1)-brane, i.e.\ an instanton-like object. The natural candidate for this are precisely the string crossing events discussed earlier. When we move the string crossing event across the domain wall we get an additional phase of $2\pi /N$, this behavior is analogous to the behavior of D(-1)- and D8-branes in type I string theory discussed by Witten in \cite{Witten:1998cd}. This suggests, that our interpretation of these objects is sound.\\
The domain walls discussed in the non-supersymmetric case seem to be more closely analogous to D8-branes in type IIA string theory. As we have seen earlier 4-form field strength $F_4$ experiences a jump upon crossing the domain wall, this is related to the case of D8-branes in type IIA, where the so called Romans mass jumps upon crossing the D8-brane \cite{Polchinski:1995mt}.

\subsection{Analogy to FQH systems}

There is one further promising possibility to extract dynamical properties. In fractional quantum Hall systems with filling factor $\nu$ one introduces a statistical gauge field with the action, see \cite{Zhang:1992eu}
\begin{equation}
S_{\text{FQH}} = \frac{i}{4 \pi \nu} \int{\mathcal{A} \wedge d \mathcal{A}} \enspace.
\end{equation}
As a consequence, for filling factor $\nu = 1/N$ we exactly recover the wall action for the (Super-)\-Yang\--Mills domain walls, \eqref{wallaction}.\\
The fundamental excitations in fractional quantum Hall systems are fractionalized electrons of charge $1/N$ with one fundamental magnetic flux attached to them. On the domain wall the magnetic field is substituted by the jump in $\theta$ and the fundamental charges on the wall are the end points of electric flux tubes of charge $1/N$. The bulk excitation on the other hand is a standard electron for the fractional quantum Hall system. The analogy for the domain walls is that $N$ fundamental electric flux tubes can end on each other and constitute a baryonic vertex.\\
The most intriguing property is that from string theory constructions we expect for walls with $\Delta k = 2$ or $\Delta \theta = 4 \pi$ the typical symmetry enhancement of coinciding D-branes. This would transform the $U(1)$ into an $U(2)$ Chern-Simons term. From our topological construction this cannot be seen because a factor of two multiplying the domain wall action would suffice to preserve the 1-form gauge invariance of the action. But this phenomenon is observed for bilayer systems in the fractional quantum Hall effect, \cite{Moore:1991ks} and \cite{Ichinose:1996nt}. There are two kinds of excitations in the bilayer systems. One transforms under a common $U(1)$ symmetry on both layers simultaneously. The other has a non-trivial structure in the labels of the layers and transforms under an additional $SU(2)$. Our hope is, that these excitations are related to flux tubes stretched between the domain walls in the gauge theories.

\section{Outlook and Conclusion}
\label{outlook}

To summarize, we succeeded in deriving an action that describes the topological properties of the confining phase of a $SU(N)$ gauge theory. This was possible with confinement generated by the condensation of charge $N$ monopoles and the transformation of charges under the Witten effect. By the possible reduction to the $\mathbb{Z}_N$ center charges instead of the full information of the $SU(N)$ gauge group a Abelian continuum description using the Abelian-Higgs model was sufficient. The resulting action describes the exchange phases of non-local operators and phases generated by string crossings in the presence of a vacuum angle. The allowed configurations are those, invariant under an additional 1-form gauge transformation.\\
Introducing domain walls in this setup while preserving the additional 1-form gauge invariance leads to the occurrence of a field on the domain wall worldvolume. This field exhibits a Chern-Simons term at level $N$ in its action reproducing the predictions of string theory constructions. Furthermore, it enables electric flux tubes to end on the domain walls. For the supersymmetric extension of the model the same mechanism is efficient.\\
Nevertheless, the dynamical properties of (non-)\-super\-symmetric walls differ. Without supersymmetry the condensate in one domain is dyonic and the corresponding 4-form field strength contributes to the energy density. In other words, the domain walls source a long-range electric field. For the supersymmetric version the phase of the gluino condensate acts as an axionic field and is eaten up by the massless 3-form, a specific form of the Higgs-mechanism. This generates a mass for the 3-form and the electric field is screened destroying the long-range interaction.\\
The analogies to string theory and the fractional quantum Hall effect furthermore offer a possible continuation of the investigation concerning dynamical properties which are not possible in our topological framework.

\begin{acknowledgments}
We would like to thank Gia Dvali, C\'esar G\'omez, Daniel Flassig and Stanislav Schmidt for fruitful discussions and useful comments.\\
A.P. was supported by the Humboldt Foundation.
\end{acknowledgments}

\bibliography{./references}

\begin{thebibliography}{31}
\expandafter\ifx\csname natexlab\endcsname\relax\def\natexlab#1{#1}\fi
\expandafter\ifx\csname bibnamefont\endcsname\relax
  \def\bibnamefont#1{#1}\fi
\expandafter\ifx\csname bibfnamefont\endcsname\relax
  \def\bibfnamefont#1{#1}\fi
\expandafter\ifx\csname citenamefont\endcsname\relax
  \def\citenamefont#1{#1}\fi
\expandafter\ifx\csname url\endcsname\relax
  \def\url#1{\texttt{#1}}\fi
\expandafter\ifx\csname urlprefix\endcsname\relax\def\urlprefix{URL }\fi
\providecommand{\bibinfo}[2]{#2}
\providecommand{\eprint}[2][]{\url{#2}}

\bibitem[{\citenamefont{Dvali and Shifman}(1997{\natexlab{a}})}]{Dvali:1996xe}
\bibinfo{author}{\bibfnamefont{G.}~\bibnamefont{Dvali}} \bibnamefont{and}
  \bibinfo{author}{\bibfnamefont{M.~A.} \bibnamefont{Shifman}},
  \bibinfo{journal}{Phys.Lett.} \textbf{\bibinfo{volume}{B396}},
  \bibinfo{pages}{64} (\bibinfo{year}{1997}{\natexlab{a}}),
  \eprint{hep-th/9612128}.

\bibitem[{\citenamefont{Witten}(1997)}]{Witten:1997ep}
\bibinfo{author}{\bibfnamefont{E.}~\bibnamefont{Witten}},
  \bibinfo{journal}{Nucl.Phys.} \textbf{\bibinfo{volume}{B507}},
  \bibinfo{pages}{658} (\bibinfo{year}{1997}), \eprint{hep-th/9706109}.

\bibitem[{\citenamefont{Dvali and Shifman}(1997{\natexlab{b}})}]{Dvali:1996bg}
\bibinfo{author}{\bibfnamefont{G.}~\bibnamefont{Dvali}} \bibnamefont{and}
  \bibinfo{author}{\bibfnamefont{M.~A.} \bibnamefont{Shifman}},
  \bibinfo{journal}{Nucl.Phys.} \textbf{\bibinfo{volume}{B504}},
  \bibinfo{pages}{127} (\bibinfo{year}{1997}{\natexlab{b}}),
  \eprint{hep-th/9611213}.

\bibitem[{\citenamefont{Acharya and Vafa}(2001)}]{Acharya:2001dz}
\bibinfo{author}{\bibfnamefont{B.~S.} \bibnamefont{Acharya}} \bibnamefont{and}
  \bibinfo{author}{\bibfnamefont{C.}~\bibnamefont{Vafa}}
  (\bibinfo{year}{2001}), \eprint{hep-th/0103011}.

\bibitem[{\citenamefont{Gaiotto}(2013)}]{Gaiotto:2013gwa}
\bibinfo{author}{\bibfnamefont{D.}~\bibnamefont{Gaiotto}}
  (\bibinfo{year}{2013}), \eprint{1306.5661}.

\bibitem[{\citenamefont{Seiberg and Witten}(1994)}]{Seiberg:1994rs}
\bibinfo{author}{\bibfnamefont{N.}~\bibnamefont{Seiberg}} \bibnamefont{and}
  \bibinfo{author}{\bibfnamefont{E.}~\bibnamefont{Witten}},
  \bibinfo{journal}{Nucl.Phys.} \textbf{\bibinfo{volume}{B426}},
  \bibinfo{pages}{19} (\bibinfo{year}{1994}), \eprint{hep-th/9407087}.

\bibitem[{\citenamefont{Witten}(1979{\natexlab{a}})}]{Witten:1978bc}
\bibinfo{author}{\bibfnamefont{E.}~\bibnamefont{Witten}},
  \bibinfo{journal}{Nucl.Phys.} \textbf{\bibinfo{volume}{B149}},
  \bibinfo{pages}{285} (\bibinfo{year}{1979}{\natexlab{a}}).

\bibitem[{\citenamefont{Witten}(1979{\natexlab{b}})}]{Witten:1979ey}
\bibinfo{author}{\bibfnamefont{E.}~\bibnamefont{Witten}},
  \bibinfo{journal}{Phys.Lett.} \textbf{\bibinfo{volume}{B86}},
  \bibinfo{pages}{283} (\bibinfo{year}{1979}{\natexlab{b}}).

\bibitem[{\citenamefont{Witten}(1998{\natexlab{a}})}]{Witten:1998uka}
\bibinfo{author}{\bibfnamefont{E.}~\bibnamefont{Witten}},
  \bibinfo{journal}{Phys.Rev.Lett.} \textbf{\bibinfo{volume}{81}},
  \bibinfo{pages}{2862} (\bibinfo{year}{1998}{\natexlab{a}}),
  \eprint{hep-th/9807109}.

\bibitem[{\citenamefont{Shifman}(1999)}]{Shifman:1998if}
\bibinfo{author}{\bibfnamefont{M.~A.} \bibnamefont{Shifman}},
  \bibinfo{journal}{Phys.Rev.} \textbf{\bibinfo{volume}{D59}},
  \bibinfo{pages}{021501} (\bibinfo{year}{1999}), \eprint{hep-th/9809184}.

\bibitem[{\citenamefont{Aharony et~al.}(2013)\citenamefont{Aharony, Seiberg,
  and Tachikawa}}]{Aharony:2013hda}
\bibinfo{author}{\bibfnamefont{O.}~\bibnamefont{Aharony}},
  \bibinfo{author}{\bibfnamefont{N.}~\bibnamefont{Seiberg}}, \bibnamefont{and}
  \bibinfo{author}{\bibfnamefont{Y.}~\bibnamefont{Tachikawa}},
  \bibinfo{journal}{JHEP} \textbf{\bibinfo{volume}{1308}}, \bibinfo{pages}{115}
  (\bibinfo{year}{2013}), \eprint{1305.0318}.

\bibitem[{\citenamefont{'t~Hooft}(1979)}]{'tHooft:1979uj}
\bibinfo{author}{\bibfnamefont{G.}~\bibnamefont{'t~Hooft}},
  \bibinfo{journal}{Nucl.Phys.} \textbf{\bibinfo{volume}{B153}},
  \bibinfo{pages}{141} (\bibinfo{year}{1979}).

\bibitem[{\citenamefont{'t~Hooft}(1981)}]{'tHooft:1981ht}
\bibinfo{author}{\bibfnamefont{G.}~\bibnamefont{'t~Hooft}},
  \bibinfo{journal}{Nucl.Phys.} \textbf{\bibinfo{volume}{B190}},
  \bibinfo{pages}{455} (\bibinfo{year}{1981}).

\bibitem[{\citenamefont{Donagi and Witten}(1996)}]{Donagi:1995cf}
\bibinfo{author}{\bibfnamefont{R.}~\bibnamefont{Donagi}} \bibnamefont{and}
  \bibinfo{author}{\bibfnamefont{E.}~\bibnamefont{Witten}},
  \bibinfo{journal}{Nucl.Phys.} \textbf{\bibinfo{volume}{B460}},
  \bibinfo{pages}{299} (\bibinfo{year}{1996}), \eprint{hep-th/9510101}.

\bibitem[{\citenamefont{Kapustin and Seiberg}(2014)}]{Kapustin:2014gua}
\bibinfo{author}{\bibfnamefont{A.}~\bibnamefont{Kapustin}} \bibnamefont{and}
  \bibinfo{author}{\bibfnamefont{N.}~\bibnamefont{Seiberg}}
  (\bibinfo{year}{2014}), \eprint{1401.0740}.

\bibitem[{\citenamefont{Gaiotto et~al.}(2010)\citenamefont{Gaiotto, Moore, and
  Neitzke}}]{Gaiotto:2010be}
\bibinfo{author}{\bibfnamefont{D.}~\bibnamefont{Gaiotto}},
  \bibinfo{author}{\bibfnamefont{G.~W.} \bibnamefont{Moore}}, \bibnamefont{and}
  \bibinfo{author}{\bibfnamefont{A.}~\bibnamefont{Neitzke}}
  (\bibinfo{year}{2010}), \eprint{1006.0146}.

\bibitem[{\citenamefont{Banks and Seiberg}(2011)}]{Banks:2010zn}
\bibinfo{author}{\bibfnamefont{T.}~\bibnamefont{Banks}} \bibnamefont{and}
  \bibinfo{author}{\bibfnamefont{N.}~\bibnamefont{Seiberg}},
  \bibinfo{journal}{Phys.Rev.} \textbf{\bibinfo{volume}{D83}},
  \bibinfo{pages}{084019} (\bibinfo{year}{2011}), \eprint{1011.5120}.

\bibitem[{\citenamefont{Gukov and Kapustin}(2013)}]{Gukov:2013zka}
\bibinfo{author}{\bibfnamefont{S.}~\bibnamefont{Gukov}} \bibnamefont{and}
  \bibinfo{author}{\bibfnamefont{A.}~\bibnamefont{Kapustin}}
  (\bibinfo{year}{2013}), \eprint{1307.4793}.

\bibitem[{\citenamefont{Horowitz}(1989)}]{Horowitz:1989ng}
\bibinfo{author}{\bibfnamefont{G.~T.} \bibnamefont{Horowitz}},
  \bibinfo{journal}{Commun.Math.Phys.} \textbf{\bibinfo{volume}{125}},
  \bibinfo{pages}{417} (\bibinfo{year}{1989}).

\bibitem[{\citenamefont{Preskill}(2000)}]{Preskill:2000}
\bibinfo{author}{\bibfnamefont{J.}~\bibnamefont{Preskill}},
  \emph{\bibinfo{title}{{Lecture Notes 230bc, Field Theory and Topology, pp.
  92-94}}} (\bibinfo{year}{2000}),
  \urlprefix\url{http://www.theory.caltech.edu/~preskill/ph230/notes2000/230Lectures7-10-Page53-109.pdf}.

\bibitem[{\citenamefont{Shifman and Vainshtein}(1988)}]{Shifman:1987ia}
\bibinfo{author}{\bibfnamefont{M.~A.} \bibnamefont{Shifman}} \bibnamefont{and}
  \bibinfo{author}{\bibfnamefont{A.}~\bibnamefont{Vainshtein}},
  \bibinfo{journal}{Nucl.Phys.} \textbf{\bibinfo{volume}{B296}},
  \bibinfo{pages}{445} (\bibinfo{year}{1988}).

\bibitem[{\citenamefont{Witten}(1982)}]{Witten:1982df}
\bibinfo{author}{\bibfnamefont{E.}~\bibnamefont{Witten}},
  \bibinfo{journal}{Nucl.Phys.} \textbf{\bibinfo{volume}{B202}},
  \bibinfo{pages}{253} (\bibinfo{year}{1982}).

\bibitem[{\citenamefont{Luscher}(1978)}]{Luscher:1978rn}
\bibinfo{author}{\bibfnamefont{M.}~\bibnamefont{Luscher}},
  \bibinfo{journal}{Phys.Lett.} \textbf{\bibinfo{volume}{B78}},
  \bibinfo{pages}{465} (\bibinfo{year}{1978}).

\bibitem[{\citenamefont{Dvali}(2005)}]{Dvali:2005an}
\bibinfo{author}{\bibfnamefont{G.}~\bibnamefont{Dvali}} (\bibinfo{year}{2005}),
  \eprint{hep-th/0507215}.

\bibitem[{\citenamefont{Dvali et~al.}(1999)\citenamefont{Dvali, Gabadadze, and
  Kakushadze}}]{Dvali:1999pk}
\bibinfo{author}{\bibfnamefont{G.}~\bibnamefont{Dvali}},
  \bibinfo{author}{\bibfnamefont{G.}~\bibnamefont{Gabadadze}},
  \bibnamefont{and}
  \bibinfo{author}{\bibfnamefont{Z.}~\bibnamefont{Kakushadze}},
  \bibinfo{journal}{Nucl.Phys.} \textbf{\bibinfo{volume}{B562}},
  \bibinfo{pages}{158} (\bibinfo{year}{1999}), \eprint{hep-th/9901032}.

\bibitem[{\citenamefont{'t~Hooft}(1974)}]{'tHooft:1973jz}
\bibinfo{author}{\bibfnamefont{G.}~\bibnamefont{'t~Hooft}},
  \bibinfo{journal}{Nucl.Phys.} \textbf{\bibinfo{volume}{B72}},
  \bibinfo{pages}{461} (\bibinfo{year}{1974}).

\bibitem[{\citenamefont{Witten}(1998{\natexlab{b}})}]{Witten:1998cd}
\bibinfo{author}{\bibfnamefont{E.}~\bibnamefont{Witten}},
  \bibinfo{journal}{JHEP} \textbf{\bibinfo{volume}{9812}}, \bibinfo{pages}{019}
  (\bibinfo{year}{1998}{\natexlab{b}}), \eprint{hep-th/9810188}.

\bibitem[{\citenamefont{Polchinski}(1995)}]{Polchinski:1995mt}
\bibinfo{author}{\bibfnamefont{J.}~\bibnamefont{Polchinski}},
  \bibinfo{journal}{Phys.Rev.Lett.} \textbf{\bibinfo{volume}{75}},
  \bibinfo{pages}{4724} (\bibinfo{year}{1995}), \eprint{hep-th/9510017}.

\bibitem[{\citenamefont{Zhang}(1992)}]{Zhang:1992eu}
\bibinfo{author}{\bibfnamefont{S.-C.} \bibnamefont{Zhang}},
  \bibinfo{journal}{Int.J.Mod.Phys.} \textbf{\bibinfo{volume}{B6}},
  \bibinfo{pages}{25} (\bibinfo{year}{1992}).

\bibitem[{\citenamefont{Moore and Read}(1991)}]{Moore:1991ks}
\bibinfo{author}{\bibfnamefont{G.~W.} \bibnamefont{Moore}} \bibnamefont{and}
  \bibinfo{author}{\bibfnamefont{N.}~\bibnamefont{Read}},
  \bibinfo{journal}{Nucl.Phys.} \textbf{\bibinfo{volume}{B360}},
  \bibinfo{pages}{362} (\bibinfo{year}{1991}).

\bibitem[{\citenamefont{Ichinose and Sekiguchi}(1997)}]{Ichinose:1996nt}
\bibinfo{author}{\bibfnamefont{I.}~\bibnamefont{Ichinose}} \bibnamefont{and}
  \bibinfo{author}{\bibfnamefont{A.}~\bibnamefont{Sekiguchi}},
  \bibinfo{journal}{Nucl.Phys.} \textbf{\bibinfo{volume}{B493}},
  \bibinfo{pages}{683} (\bibinfo{year}{1997}), \eprint{cond-mat/9610054}.

\end{thebibliography}
\end{document}